\documentclass[a4paper,11pt]{article}
\pdfoutput=1 % if your are submitting a pdflatex (i.e. if you have
\usepackage{jheppub} % for details on the use of the package, please
\usepackage{graphicx,amssymb,bm,latexsym,amsmath}
\usepackage{epstopdf}
\usepackage{caption}
\usepackage{subcaption}
\usepackage[section]{placeins}
\usepackage{float}
\usepackage{capt-of}
\usepackage{tabu}

\newcommand{\mathsym}[1]{{}}
\newcommand{\unicode}[1]{{}}

\title{\boldmath Spinning pulsating strings in $(AdS_5 \times S^5)_{\varkappa}$ }

\author{Sorna Prava Barik,}
\author{Kamal L. Panigrahi,}
\author{Manoranjan Samal}
\affiliation{Department of Physics,\\Indian Institute of Technology Kharagpur,\\
	Kharagpur-721 302, India}

\emailAdd{sorna}
\emailAdd{panigrahi}
\emailAdd{manoranjan@phy.iitkgp.ernet.in}

\abstract{ We study a general class of spinning pulsating strings in $(AdS_5 \times S^5)_{\varkappa}$ background. For these family of solitons, we examine the scaling relation between the energy, spin or angular momentum. We verify that in $\varkappa \rightarrow 0 $ limit these relations reduce to the undeformed $AdS_5 \times S^5$ case. We further study an example of a string which is spinning in the $\varkappa$-deformed AdS$_5$ and S$^5$ simultaneously and find out the scaling relation among various conserved charges.}

\begin{document}
	\maketitle
	\flushbottom
	\section{Introduction}
	The AdS/CFT correspondence relates string states on AdS and gauge invariant operators in the gauge theory side. The most studied example of the AdS/CFT duality is the one between spectrum of closed superstrings (supergravity) in $AdS_5 \times S^5$ background and gauge invariant operators in four dimensional $\mathcal{N} = 4$ Supersymmetric Yang-Mills (SYM) theory based on the gauge group $SU(N)$ \cite{Maldacena:1997re, Witten:1998qj, Aharony:1999ti}.  A remarkable feature of AdS/CFT duality is an underlying integrability \cite{Bena:2003wd} structure on both side, which provides an important tool for finding the spectrum on both sides and many significant properties have been revealed based on exact computations. To understand the structure of full string spectrum, one need to identify the classical solitonic solutions of $AdS_5\times S^5$ sigma model carrying global charges. The Yang-Mills theory itself can be mapped to a integrable spin-chain system \cite{Minahan:2002ve}. The basic idea of relating all string states to precise dual gauge theory operators is a tough job due to presence of infinite tower of string solutions on string theory side. One probable way out of this problem is that in the large angular momentum or large R-charge limit both sides of the duality become more tractable.
	One of the advantages of this limit is that the anomalous dimension of operators in the SYM theory can be related to the dispersion relation between conserved charges of spinnings and pulsating strings in the large charge limit. 
	
	In this context, a large variety of rotating and spinning strings has been studied in $AdS_5 \times S^5$ precisely and also have been mapped to dual spin-chain excitations. 
	These include the already well studied giant magnon, folded strings and spiky strings solutions and the gauge theory duals have been analyzed in great detail. In spinning string case the highly excited string states corresponds to the gauge theory operators with small anomalous dimension. This type of  strings is the generalization of the folded \cite{Gubser:2002tv}, \cite{deVega:1996mv} and spiky \cite{Kruczenski:2004wg} strings with single spin in AdS$_3$ part of AdS$_5$. The semiclassical multi spinning string states (strings spinning in AdS$_5$ ) have also been found to be dual to  certain trace operators \cite{Tirziu:2009ed}. It has been shown that for these solutions, the string states are unstable  for large charges \cite{Frolov:2003qc}. On the other hand, the circular pulsating string solutions have been less explored. Pulsating strings were first introduced in \cite{Minahan:2002rc} where they were expected to correspond to certain highly excited sigma model
	operators and later on were generalized to  \cite{Engquist:2003rn}, \cite{Dimov:2004xi},  \cite{Smedback:1998yn}. In \cite{Minahan:2002rc} and  \cite{Beccaria:2010zn}, pulsating string solutions in $AdS_5$ and $S^5$ respectively have been worked out separately where as in \cite{Park:2005kt}, a string rotating and at the same time oscillating in $AdS_5$ have been derived. It has been shown that the addition of oscillation to spinning strings improve the stability of the string states \cite{Khan:2005fc}.  An interesting class of solutions were proposed in \cite{Khan:2003sm} which generalized some of the earlier pulsating and spinning string solutions by looking at strings which are straight and spinning in one direction but circular and pulsating in another, and with a non-trivial coupling between the two in $AdS_5 \times S^5$ background. Even though the exact gauge theory operators corresponding to these class of string states are still unknown, they are interesting in their own right. In this paper, we wish to generalize such spinning pulsating strings in $(AdS_5 \times S^5)_{\varkappa}$ background, the sigma model associated to which retains integrability of the original model. 
	
	To explore the integrability beyond the usual $AdS_5 \times S^5$, it is natural to consider the integrable deformation of the background to study the string motion as well as the underlying gauge theory, if any. Following the earlier proposals about Yang-Baxter deformation made in \cite{Klimcik:2002zj,Klimcik:2008eq,Klimcik:2014bta}, a one-parameter deformed $AdS_5 \times S^5$ supercoset model was constructed in  \cite{Delduc:2013qra,Delduc:2014kha,Hoare:2014pna}. The background string metric and the NS-NS 2-form corresponding to deformed structure has been worked out \cite{Arutyunov:2013ega}. The symmetry group $SO(2, 4)\times SO(6)$ of $AdS_5 \times S^5$  reduced to its Cartan subgroup $[U(1)]^6$ and it produced a nice laboratory to study string motion. 
	Given the integrable nature of the deformed background, various rigidly rotating and pulsating strings have been investigated in detail \cite{Khouchen:2014kaa},\cite{Banerjee:2014bca},\cite{Panigrahi:2014sia}, \cite{Arutynov:2014ota}, \cite{Hernandez:2017raj},\cite{Banerjee:2016xbb}. Since the background suffers from the presence of a singularity surface in the AdS space, a new coordinate system to handle this has been developed \cite{Kameyama:2014vma}. The folded GKP like string solutions of   \cite{Kameyama:2014vma} generalized to N-spike string solutions in  \cite{Banerjee:2015nha}. In these cases, however, it was found out that in the ‘long’ string limit, where the strings touch the singularity surface, the expression for cusp anomalous dimension does not reduce to the undeformed one in the $\varkappa \rightarrow 0$ limit and a possible explanation was provided in 
	\cite{Kameyama:2014vma}. Not only classical string solutions, various minimal surfaces and Wilson loops in this background have been found in \cite{Kameyama:2014via}. In some recent works \cite{Araujo:2017jap},\cite{Araujo:2017jkb}, conformal twist and non-commutative gauge theory have been studied in Yang-Baxter deformed AdS$_5 \times $ S$^5$ background.
	
	In the present paper, we are interested in a non-trivial string solitons in deformed $AdS_5 \times S^5$ background with $\varkappa$ being the deformed parameter. The strings, we consider here
	are straight and spinning in one sub-space while it is circular and pulsating in another. In each case, we obtain the exact form of string solutions in terms of elliptic functions and discuss their properties. We construct the scaling relations between energy and charges in various limits and compare the result with the undeformed results. We would like to mention in passing that the dual gauge theory of the $\varkappa$-deformed background is not yet known and hence the operator dual to these class of solutions is not apriori clear. One of the main reasons is the lack of our understanding of incorporating the deformed parameter into the operators which would be present in the unknown gauge theory. Nevertheless studying the string states will surely enhance our understanding of the string theory in integrable deformed background. The rest of the paper is organized as follows. In section-2, we outline the background that we are interested in for studying the string solutions. In section-3, we study the strings which are pulsating in AdS$_5$ subspace and is spinning in $S^5$ simultaneously. We find out the scaling relation among the energy and the angular momentum. We study the behavior of the long and short strings in this case and have checked that they indeed reduce to the similar relations in the undeformed $AdS_5 \times S^5$ case. Section-4 is devoted to the study of strings which pulsate in S$^5$ and at the same time spins along the deformed AdS$_5$. We have looked at some interesting spinning string solution which spins along both deformed $AdS_5$ and $S^5$ subspace and write down the energy spin relationship in section-5. In section-6 we conclude with some discussions.

	\section{The $\varkappa  $  deformed $ \left(\text{AdS}_5 \times \text{S}^5\right)_\varkappa $ geometry }
	The so called  $\varkappa$ deformed $ \left(\text{AdS}_5 \times \text{S}^5\right)_\varkappa $ background is
	\begin{eqnarray}
	ds^2_{AdS_5}&=&-h(\rho) dt^2+ f(\rho) d\rho^2+\rho^2 \left(u(\rho,\psi)(d \psi^2+\cos^2 \psi d \psi^2_1)+\sin^2 \psi d\psi_2^2\right), \\
	ds^2_{S^5}&=& \tilde{h}(r) d \varphi^2+ \tilde{f}(r)dr^2+r^2\left(\tilde{u}(r,\phi)(d\phi^2+\cos^2 \phi d \phi^2_1)+\sin^2 \phi d\phi_2^2\right),
	\end{eqnarray}
	where $$h(\rho)=\frac{1+\rho^2}{1-\varkappa^2 \rho^2},~~~~f(\rho)=\frac{1}{(1+\rho^2)(1-\varkappa^2 \rho^2)},~~~~u(\rho,\psi)=\frac{1}{1+\varkappa^2 \rho^4 \sin^2 \psi},$$
	and  $$\tilde{h}(r)=\frac{1-r^2}{1+\varkappa^2 r^2},~~~~\tilde{f}(r)=\frac{1}{(1-r^2)(1+\varkappa^2 r^2)},~~~~ \tilde{u}(r,\phi)=\frac{1}{1+\varkappa^2 r^4 \sin^2 \phi},$$
	supported by the following NS-NS B field
	\begin{equation}
	B_{\psi_1 \psi}=\frac{1}{2}\varkappa^2 \rho^4 \sin 2 \psi ~u(\rho,\psi),~~~~~~~
	B_{\phi_1 \phi}=-\frac{1}{2} \varkappa^2 r^4 \sin 2 \phi ~ \tilde{u}(r,\phi),
	\end{equation}
	where $\varkappa \in [0,\infty)$ is the deforming parameter. It can be seen that for $\varkappa=0$, the above geometry reduces to the original $ \text{AdS}_5 \times \text{S}^5 $. It is convenient to write the above metric and the NS-NS B-field components in global coordinates which is obtained by the following  co-ordinate transformations: $\rho \rightarrow \sinh \rho$ and $r \rightarrow \cos \theta $. The metric in the global coordinates looks like   
	\begin{eqnarray}
	ds^2_{AdS^5}&=&-\frac{\cosh^2 \rho}{1-\varkappa^2 \sinh^2 \rho}dt^2+\frac{d\rho^2 }{1-\varkappa^2 \sinh^2 \rho}+\frac{\sinh^2 \rho d \psi^2}{1+\varkappa^2 \sinh^4 \rho \sin^2 \psi} \nonumber \\ &~& \hspace{3cm}+ \frac{\sinh^2 \rho \cos^2 \psi d \psi_1^2}{1+\varkappa^2 \sinh^4 \rho\sin^2 \psi}+\sinh^2 \rho \sin^2 \psi d\psi_2^2, \\
	ds^2_{S^5} &=& \frac{\sin^2 \theta d\varphi^2}{1+\varkappa^2 \cos^2 \theta}+\frac{d \theta^2}{1+\varkappa^2 \cos^2 \theta}+\frac{\cos^2 \theta d \phi^2}{1+\varkappa^2 \cos^4 \theta \sin^2 \phi}  \nonumber \\  
	&~& \hspace{3cm}+\frac{\cos^2 \theta \cos^2 \phi d \phi_1^2}{1+\varkappa^2 \cos^4 \theta \sin^2 \phi}+\cos^2 \theta \sin^2 \phi d\phi_2^2.
	\end{eqnarray}
	We would like to study a general class of spinning pulsating string in this background.
	The deformed bosonic string action in conformal gauge can be written as
	\begin{equation}
	S=\hat{T} \int d^2 \sigma \left(\sqrt{-h}h^{\alpha \beta} G_{\mu \nu}\partial_ \alpha 	X^\mu  \partial_ \beta X^\nu+\epsilon^{\alpha \beta}B_{\mu \nu}\partial_ \alpha	X^\mu  \partial_ \beta X^\nu \right),
	\end{equation}
	where $\hat{T}=\frac{1}{4 \pi \alpha'}\sqrt{1+\varkappa^2}$ is the effective string tension.
	\section{Pulsating in $\left(\text{AdS}_5\right)_\varkappa$ and spinning in $\left(\text{S}^5\right)_\varkappa$ }
	In this section, we consider a string which is spinning in $\text{S}^5$ and pulsating in  $\text{AdS}_5$ of the deformed background. We use the following ansatz 
	\begin{eqnarray}
	t=t(\tau),~~\rho=\rho(\tau),~~ \psi=\frac{\pi}{2},~~\psi_2 =\sigma, \nonumber \\
	\varphi= \varphi (\tau),~~ \theta= \theta(\sigma),~~\phi=\frac{\pi}{2},~ \phi_2 =const.
	\end{eqnarray}
	Taking the above ansatz the Polyakov action in conformal gauge takes the form,
	\begin{equation}
	S=\hat{T}\int d^2 \sigma \left(-\frac{\cosh^2 \rho\dot{t}^2}{1-\varkappa^2 \sinh^2 \rho}+\frac{\dot{\rho}^2}{1-\varkappa^2 \sinh^2 \rho}-\sinh^2 \rho +\frac{\sin^2 \theta}{1+\varkappa^2 \cos^2 \theta}\dot{\varphi}^2-\frac{\theta'^2}{1+\varkappa^2 \cos^2 \theta}\right).
	\end{equation}
	We  notice that for such ansatz the  B-fields do not contribute to the action.
	The equation of motion for $t$ and $\varphi $ can be found as
	\begin{equation} \label{spinning in s eom-t}
	\dot{t}= \frac{c_0 (1-\varkappa^2 \sinh^2 \rho)}{\cosh^2 \rho},~~~ \dot{\varphi}= \nu, 
	\end{equation}
	where $c_0$ and $\nu$ are the integration constants. From the equation motion of the $\varphi$, we find that the string rotates in $\varphi$ direction with angular velocity $\nu$. Now, the equations of motion for $\rho$ and $\theta$ can be written as
	\begin{equation}
	2 \ddot{ \rho}(1-\varkappa^2 \sinh^2 \rho)+\varkappa^2 \sinh^2 \rho \dot{\rho}^2+\sinh^2 2 \rho\left((1+\varkappa^2)\dot{t}^2+(1-\varkappa^2 \sinh^2 \rho)^2\right)=0,
	\end{equation}
	and 
	\begin{equation}
	2 \theta''\left(1+\varkappa^2 \cos^2 \theta \right)+\varkappa^2\theta'^2\sin 2 \theta +\nu^2\sin 2 \theta (1+\varkappa^2)=0.
	\end{equation}
	These equations of motion are to be supported by the conformal gauge constraints
	\begin{eqnarray}
	G_{m n}\left( \partial _{\tau} X^m \partial _{\tau} X^n +\partial _{\sigma} X^m \partial _{\sigma} X^n \right) =0 \label{constraint1} \\
	G_{mn}\partial _{\tau} X^m \partial _{\sigma} X^n=0.
	\end{eqnarray}
	From (\ref{constraint1}), we get
	\begin{equation}
	-\frac{c_0^2(1-\varkappa^2 \sinh^2 \rho)}{\cosh^2 \rho}+\frac{\dot{\rho}^2}{1-\varkappa^2 \sinh^2 \rho}+\sinh^2 \rho+\frac{\nu^2 \sin^2 \theta}{1+\varkappa^2 \cos^2 \theta}+\frac{\theta'^2}{1+\varkappa^2 \cos^2 \theta}=0.
	\end{equation}
	The above equation is a coupled equation of $\rho(\tau)$ and $\theta(\sigma)$. However without loss of generality we can separate the above equation as
	\begin{eqnarray}
	-\frac{c_0^2(1-\varkappa^2 \sinh^2 \rho)}{\cosh^2 \rho}+\frac{\dot{\rho}^2}{1-\varkappa^2 \sinh^2 \rho}+\sinh^2 \rho+c_1^2=0, \\
	\frac{\nu^2 \sin^2 \theta}{1+\varkappa^2 \cos^2 \theta}+\frac{\theta'^2}{1+\varkappa^2 \cos^2 \theta}-c_1^2=0,
	\end{eqnarray}
	where $c_1$ is a constant. One can check that the above equations are consistent with the equations of motion $\rho$ and $\theta$. Now solving these equations we find the following solution of $\rho$ and $\theta$ 
	\begin{equation}
	\sinh \rho=\sqrt{\frac{R_- \mathbf{sn}^2\left((1+\varkappa^2 c_1^2) \sqrt{R_+}\tau\bigg\rvert\frac{R_-}{R_+}\right)}{1+\varkappa^2\left[1+R_- \mathbf{sn}^2\left((1+\varkappa^2 c_1^2) \sqrt{R_+}\tau\bigg\rvert\frac{R_-}{R_+}\right)\right]}},
	\end{equation}
	\begin{equation}
	\sin\theta=\sqrt{\frac{c_1^2(1+\varkappa^2)}{\nu^2+\varkappa^2c_1^2}} \mathbf{sn}\left(\sqrt{\nu^2+\varkappa^2 c_1^2}\sigma\bigg\rvert\frac{c_1^2(1+\varkappa^2)}{\nu^2+\varkappa^2c_1^2}\right),
	\end{equation}
	with \begin{equation}\label{roots of theta}
	R_\pm=\frac{-1+\varkappa^2 c_0^2-c_1^2(1+2\varkappa^2)\pm\sqrt{(c_1^2-1)^2+\varkappa^4c_0^4+2c_0^2(2+(1+c_1^2)\varkappa^2)}}{2+2\varkappa^2c_1^2}.
	\end{equation}
	Now we take periodicity condition of $\theta$ which gives, the range $0\leq \sigma \leq 2 \pi$ is divided into four equal segments: in the first segment i.e. for $0\leq \sigma \leq \frac{\pi}{2}$, the solution of $\theta(\sigma)$ increases from zero to its maximal value $\theta_{max}=\sin^{-1}\sqrt{\frac{c_1^2(1+\varkappa^2)}{\nu^2+\varkappa^2c_1^2}}$, then it decreases to zero in the $\frac{\pi}{2}\leq \sigma \leq \pi$ segment and continues the similar pattern for $\pi\leq \sigma \leq 2 \pi$ region.
	\begin{eqnarray}
	2\pi &=& \int_0^ {2\pi}  d \sigma= 4\int_0^{\theta_{max}} \frac{d \theta}{\theta'} =4\int_0^{\theta_{max}} \frac{d \theta}{\sqrt{c_1^2(1+\varkappa^2)-\sin^2 \theta (\nu^2+ c_1^2 \varkappa^2)}}.
	\end{eqnarray}
	Using the definition of complete elliptic integral of first kind ($\mathbf{K}$), we get 
	
	\begin{equation} \label{spinning in S5 periodicity condition  }
	2\pi \sqrt{\nu^2 +\varkappa^2 c_1^2}=4 \mathbf{K}\left[\frac{c_1^2(1+\varkappa^2)}{\nu^2+\varkappa^2c_1^2}\right].
	\end{equation}
	The conserved charges can be evaluated as
	\begin{eqnarray}
	E &=& \int_0^{2 \pi} \frac{\sqrt{1+\varkappa^2} c_0}{2 \pi \alpha' } d \sigma  =\frac{c_0}{\alpha'} \sqrt{1+\varkappa^2},\\ \label{pulsating in ads enegry}
	\mathcal{J} &=& \frac{\nu \sqrt{1+\varkappa^2}}{2 \pi \alpha' } \int_0^{2 \pi} \frac{\sin^2 \theta}{1+\varkappa^2 \cos^2 \theta} d \sigma. \label{pulsating in ads J}
	\end{eqnarray}
	Let us define two parameters $A=\sinh \rho_{max}=\sqrt{R_-}$  and  $B=\sin \theta_{max}=\sqrt{\frac{c_1^2(1+\varkappa^2)}{\nu^2+\varkappa^2c_1^2}}$. Now we want to express the conserved charges in terms of these two parameters so that we can analyze both ``short string" and ``long string" limits. Let us first find out the other parameters in terms  $A$ and $B$. Using the periodicity condition (\ref{spinning in S5 periodicity condition  }), we get
	\begin{equation}\label{parameters in A and B}
	c_1=\frac{2 B \mathbf{K}(B^2)}{\pi \sqrt{1+\varkappa^2}} \text{~~and~~} \nu=\frac{2}{\pi}\sqrt{\frac{1+\varkappa^2(1-B^2)}{1+\varkappa^2}}\mathbf{K}(B^2).
	\end{equation}
	Solving (\ref{roots of theta}) for $c_0$ and then substituting the value of $c_1$ from the above equation, we get the conserved energy in the following form
	\begin{equation}
	E=\sqrt{\frac{(1+A^2)\left[ (1+\varkappa^2+  \varkappa^2 A^2)4 B^2 \mathbf{K}^2(B^2)+\pi^2 A^2 (1+\varkappa^2) \right]}{\alpha'^2\pi^2 \left(1+\varkappa^2(1+A^2)\right)}}.
	\end{equation} 
	Similarly substituting the value of $\sin \theta$ in (\ref{pulsating in ads J}) and then using (\ref{parameters in A and B}), we find the expression of angular momentum in terms $B$ as
	\begin{eqnarray}
	\mathcal{J}=\frac{2\sqrt{1+\varkappa^2-\varkappa^2B^2}}{\alpha'\pi \varkappa^2}\left[\mathbf{\Pi}\left(\frac{\varkappa^2 B^2}{1+\varkappa^2},B^2\right)-\mathbf{K}(B^2)\right].
	\end{eqnarray}
	%\begin{equation}
	%\mathcal{J}=\frac{2\sqrt{1+\varkappa^2-\varkappa^2B^2}}{\pi  (1+\varkappa^2)}\left[\mathbf{K}(B^2)-\mathbf{E}(B^2)-\frac{\varkappa^2}{3(1+\varkappa^2)}\left(2(1+B^2) \mathbf{E}(B^2)-(2+B^2 )\mathbf{K}(B^2)\right)\right]. 
	%\end{equation}\par
	Now we reach in a position where we can apply short string limit or long string limit to  establish the relation between conserved charges. Let us discuss these two situations one by one.\\ 
	\textbf{\underline{Short string limit}} \\
	Consider the short-string solution in $\text{S}^5$  which correspond to  $\sin\theta_{max}<<1$ or $B<<1$. For which the leading order term in energy and angular momentum turn out to be
	\begin{equation}
	E\approx\sqrt{\frac{(1+A^2)\left[ (1+\varkappa^2+  \varkappa^2 A^2) B^2 + A^2 (1+\varkappa^2) \right]}{\alpha'^2 \left(1+\varkappa^2(1+A^2)\right)}},
	\end{equation}
	\begin{equation}
	\mathcal{J}\approx\frac{B^2}{2 \alpha'\sqrt{1+\varkappa^2}}.
	\end{equation}
	Therefore the scaling relation between $E$ and $\mathcal{J}$ becomes
	\begin{equation}
	E\approx\sqrt{\frac{(1+A^2)\left[ 2\alpha'(1+\varkappa^2+  \varkappa^2 A^2)  \sqrt{1+\varkappa^2}\mathcal{J} + A^2 (1+\varkappa^2) \right]}{\alpha'^2 \left(1+\varkappa^2(1+A^2)\right)}}.
	\end{equation}
	Now let us consider two sub cases depending upon the nature of the oscillation of string in $AdS_5$ part. For $A<<1$ which corresponds to small oscillation at center AdS, the relation between $E$ and $\mathcal{J}$ becomes
	\begin{equation}
	\alpha' E^2\approx 2 J \sqrt{1+\varkappa^2}.
	\end{equation}
	For $\varkappa \rightarrow 0$, we get $\alpha' E^2 \approx 2 \mathcal{J}$, which is a well known result for undeformed $AdS_5 \times S^5$ as found in\cite{Khan:2003sm},\cite{Russo:2002sr}.
	Similarly, for $A>>1$, we write the scaling relation by expanding it for small $\varkappa$
	\begin{equation}
	E=\left(\frac{A^2}{\alpha'}+\frac{1}{2\alpha'}+\mathcal{J}\right)+\left(-\frac{A^4}{2 \alpha'}+\left(\mathcal{J}-\frac{1}{2 \alpha'}\right)\frac{A^2}{2}+\left(\frac{3\mathcal{J}}{4}+\frac{1}{16 \alpha'}\right)\right)\varkappa^2+\mathcal{O}(\varkappa^4).
	\end{equation}
	We can notice, as expected, the zeroth order $\varkappa$ term matches with its corresponding  undeformed case as found in \cite{Khan:2003sm}.\\
	\textbf{\underline{Long string limit}} \\
	For long string in $S^5$ , the string extends to the equator so  $B\approx1$ for which $E$ and $\mathcal{J}$ become
	\begin{equation}
	E\approx\sqrt{\frac{(1+A^2)\left((1+\varkappa^2+\varkappa^2 A^2)\log^2\frac{16}{1-B^2}+  \pi^2 (1+\varkappa^2)A^2 \right)}{\alpha'^2 \pi^2 \left(1+\varkappa^2(1+A^2)\right)}},
	\end{equation}
	\begin{equation}
	\mathcal{J}\approx \frac{1}{\alpha'\pi}\log \frac{16}{1-B^2}-\frac{6+2 \varkappa^2}{3 \pi \alpha'}.
	\end{equation}
	Therefore the relation between $E$ and $\mathcal{J}$ turns out to be
	\begin{equation}
	E\approx\sqrt{\frac{(1+A^2)\left[\left(1+\varkappa^2+\varkappa^2A^2\right) \left(\alpha'\pi\mathcal{J}+2 +\frac{2}{3} \varkappa^2\right)^2+\pi^2 (1+\varkappa^2)A^2\right]}{\alpha'^2 \pi^2 \left[1+\varkappa^2(1+A^2)\right]}}.
	\end{equation} 
	For small oscillation in $AdS$ i.e $A <<1$, we get
	\begin{equation}
	E-\mathcal{J}\approx \frac{2}{\pi \alpha'}\left(1+\frac{\varkappa^2}{3}\right)+\frac{\mathcal{J}A^2}{2}.
	\end{equation}
	On the other hand, for $A>>1$ we have to distinguish two different scenarios \\
	(i) for $\alpha'J<<A$, we get the scaling relation as
	\begin{equation}
	E=\left(\frac{A^2}{\alpha'}+\frac{\alpha' \mathcal{J}^2}{2}\right)+\left(-\frac{A^4}{2 \alpha'}+\left(\frac{\mathcal{J}}{\alpha'}+\frac{ \alpha'\mathcal{J}^2}{4}\right)\frac{A^2}{2}\right)\varkappa^2+\mathcal{O}(\varkappa^4).
	\end{equation}
	(ii) for  $\alpha'J>>A$, we get
	\begin{equation}
	E=A \mathcal{J}+\frac{A^3}{2 \alpha'^2 \mathcal{J}}+\left(\frac{2 A^2}{3 \pi \alpha'}-\frac{A^3}{3 \pi \alpha'^3 \mathcal{J}^2}+\left(-\frac{1}{2 \alpha'^2 \mathcal{J}}+\frac{1}{\alpha'^3 \pi \mathcal{J}^2}\right)A^5\right)\varkappa^2+\mathcal{O}(\varkappa^4).
	\end{equation}
	We can notice that for $\varkappa\rightarrow 0$ the above results agree with their undeformed case  as found in \cite{Gubser:2002tv}, \cite{Khan:2003sm},\cite{Russo:2002sr}.
	\section{Spinning in $\left(\text{AdS}_5\right)_\varkappa$ and Pulsating in  $\left(\text{S}^5\right)_\varkappa$}
	In this section, we consider a string which is spinning in $\text{AdS}_5$ and pulsating in $\text{S}^5$ of the deformed geometry. The corresponding string ansatz:
	\begin{equation}
	t=k \tau, ~~~\rho= \rho(\sigma), ~~~  \psi= \frac{\pi}{2},~~~\psi_2= \omega \tau,~~~ \theta=\theta(\tau),~~~ \varphi= \sigma.
	\end{equation}
	The Polyakov action for the above string ansatz can be written as
	\begin{equation}
	S_p=\hat{T} \int d^2 \sigma \left(-\frac{k^2\cosh^2 \rho  }{1-\varkappa^2 \sinh^2 \rho }- \frac{ \rho'^2}{1-\varkappa^2 \sinh^2 \rho}+\omega^2\sinh^2\rho+ \frac{ \dot{\theta}^2}{1+\varkappa^2 \cos^2 \theta}-\frac{\sin^2 \theta}{1+\varkappa^2 \cos^2 \theta }  \right).
	\end{equation}
	The equations of motion for $\rho$ and $\theta$ can be found as
	\begin{eqnarray}
	2 \rho''(1-\varkappa^2 \sinh^2 \rho)+\varkappa^2 \rho'^2\sinh 2 \rho+\sinh 2 \rho\left(\omega^2(1-\varkappa^2 \sinh^2 \rho)^2-k^2(1+\varkappa^2)\right)=0, \label{spinning in AdS EOM of rho}\\
	2 \ddot{\theta}(1+\varkappa^2 \cos ^2 \theta)+\dot{\theta}^2 \varkappa^2 \sin 2 \theta+(1+\varkappa^2) \sin2 \theta=0. \label{spinning in AdS EOM of theta}
	\end{eqnarray}
	The above equations of motion are supplemented with the following conformal gauge constraints:
	\begin{equation}
	-\frac{\cosh^2 \rho k^2 }{1-\varkappa^2 \sinh^2 \rho }+ \frac{ \rho'^2}{1-\varkappa^2 \sinh^2 \rho}+\omega^2\sinh^2\rho+ \frac{ \dot{\theta}^2}{1+\varkappa^2 \cos^2 \theta}+\frac{\sin^2 \theta}{1+\varkappa^2 \cos^2 \theta }=0. 
	\end{equation}
	We can decouple the equations of $\rho$ and $\theta$ as done in the previous section.  
	\begin{equation}\label{spinning in ads constraint for rho}
	\frac{-\cosh^2 \rho  k^2}{1-\varkappa^2 \sinh^2\rho} + \frac{\rho'^2}{1-\varkappa^2 \sinh^2 \rho}+ \omega^2 \sinh^2 \rho + c_1^2=0,
	\end{equation}
	\begin{equation}\label{spinning in ads constraint for theta}
	\frac{\dot{\theta}^2}{1+\varkappa^2 \cos^2 \theta}+ \frac{ \sin^2\theta}{1+\varkappa^ 2\cos^2 \theta}-c_1^2=0,
	\end{equation}
	where $c_1$ is a constant. We can also check that the above equations are consistent with equations of motion of $\rho$ (\ref{spinning in AdS EOM of rho}) and $\theta$ (\ref{spinning in AdS EOM of theta}). Now it is easy to obtain the solution of $\rho$ and $\theta$ from these equations. From (\ref{spinning in ads constraint for rho}) we get
	\begin{equation}\label{roots of rho}
	\rho'^2=\varkappa^2 \omega^2(\sinh^2 \rho-\sinh^2 \rho_-)(\sinh^2 \rho-\sinh^2 \rho_+),
	\end{equation}
	where 
	\begin{equation}
	z_{\pm}=\sinh^2 \rho_{\pm}= \frac{\omega^2 -k^2 - c_1^2 \varkappa^2\pm\sqrt{(\omega^2 -k^2 -c_1^2 \varkappa^2)^2- 4 \varkappa^2 \omega^2(k^2-c_1^2)}}{2 \varkappa^2 \omega^2}.
	\end{equation} 
	Now integrating (\ref{roots of rho}), we get the following solution of $\rho$ in terms of elliptic function
	\begin{equation}
	\coth^2\rho=\frac{z_++1}{z_+}\frac{\mathbf{dc}^2\left(\varkappa \omega\sqrt{z_+(1+z_-)}\sigma,\mathbf{k}^2\right)}{\mathbf{k}^2},
	\end{equation}
	where \textbf{dc} is Jacobi elliptic \textbf{dc} function and $\mathbf{k}^2= \frac{z_-(1+z_+)}{z_+(1+z_-)}
	$. Similarly the solution of $\theta$ can be found from (\ref{spinning in ads constraint for theta}) as
	
	\begin{equation}
	\sin \theta= 
	\begin{cases}
	\mathbf{sn} \left(c_1 \sqrt{1+\varkappa^2} \tau, \frac{1}{B^2}
	\right) 		& \text{for~} B > 1 \\
	B~\mathbf{sn}\left( \sqrt{1+c_1^2 \varkappa^2} \tau , B^2
	\right)            & \text{for~} B < 1,
	\end{cases}
	\end{equation}
	where 
	$B=\frac{c_1\sqrt{1+\varkappa^2}}{\sqrt{1+c_1^2 \varkappa^2}}$ . The above  solution  of $\theta$ implies  the pulsating nature of string along $\theta$ direction.
	
	The positiveness of the left hand side of (\ref{roots of rho}) put constraint on the values of $\rho$, which gives the allowed regions of $\rho$ are
	\begin{equation}\label{accessible region of rho}
	(i)~~	0 ~\leq~ \rho ~\leq~ \rho_- ~~~~~~~~\text{with~~} \rho_-\leq \rho^* < \rho_s 
	\end{equation}
	\begin{equation}
	(ii)~~	\rho_+ ~\leq~ \rho < ~\infty ~~~~~~ \text{with~~} \rho^*\leq \rho_+< \rho_s
	\end{equation}
	where $\rho^*$ is the value of $\rho$ where $\rho_-=\rho_+$ and $\rho_s=\sinh^{-1}\left(\frac{1}{\varkappa}\right)$ is the singularity in $\left(AdS_5\right)_\varkappa$
	In our case i.e for spinning string solution we study the string motion in  the first region because it contains the center of AdS. Again since $\sinh^2 \rho_{\pm} \in \mathbb{R}_+$ it gives the following  inequality 
	\begin{equation} \label{inequality condition }
	\omega \geq \left(\frac{\sqrt{1+\varkappa^2}+\varkappa\sqrt{1-\frac{c_1^2}{\omega^2}}}{1+\frac{c_1^2}{\omega^2} \varkappa^2}\right)k.
	\end{equation} 
	The periodicity condition can be written as 
	\begin{equation*}
	2 \pi=\int_0^{2 \pi} d \sigma= 4\int_0^{\rho_-} \frac{d \rho}{\rho'},
	\end{equation*}
	which gives the following relation between the parameters
	\begin{equation}\label{periodicity condition}
	\pi \varkappa \omega \sqrt{z_+(z_-+1)}=2 \mathbf{K}\left(\frac{z_-(z_++1)}{z_+(z_-+1)
	}\right).
	\end{equation}
	The conserved energy and spin can be evaluated as
	\begin{eqnarray}
	E&=&\frac{k\sqrt{1+\varkappa^2}}{2 \pi \alpha'} \int_0^{2 \pi}\frac{\cosh^2 \rho}{1-\varkappa^2 \sinh^2 \rho } d \sigma=\frac{k\sqrt{1+\varkappa^2}}{2 \pi \alpha'} \int_0^{2 \pi}\frac{\cosh^2 \rho}{1-\varkappa^2 \sinh^2 \rho } \frac{d \rho}{\rho'}, \nonumber \\
	S &=&\frac{\omega\sqrt{1+\varkappa^2}}{2 \pi \alpha'}  \int_0^ {2 \pi} \sinh^2 \rho d \sigma= \frac{\omega\sqrt{1+\varkappa^2}}{2 \pi \alpha'}  \int_0^ {2 \pi} \sinh^2 \rho\frac{d \rho}{\rho'}.
	\end{eqnarray}
	Substituting the value of $\rho'$ from (\ref{roots of rho}), we get
	\begin{eqnarray}
	E&=&\frac{2 k \sqrt{1+\varkappa^2}}{\pi \alpha'\varkappa \omega \sqrt{z_+(z_-+1)}} \mathbf{\Pi}\left(\frac{z_-(1+\varkappa^2)}{z_-+1},\frac{z_-(z_++1)}{z_+(z_-+1)	}
	\right), \label{spinning in ads general energy} \\
	S&=& \frac{2\sqrt{ 1+\varkappa^2}}{\pi \alpha'\varkappa\sqrt{z_+(z_-+1)}}
	\left( \mathbf{\Pi}\left(\frac{z_-}{z_-+1},\frac{z_-(z_++1)}{z_+(z_-+1)}
	\right)- \mathbf{K
	}\left(\frac{z_-(z_++1)}{z_+(z_-+1)}\right) 
	\right),	\label{spinning in ads general spin}
	\end{eqnarray}
	\underline{\textbf{Short string limit}}\\
	As the maximum value of $\rho$ is $\rho_-$,  for short string limit $\rho_- \rightarrow 0$ or $z_- \rightarrow 0 $. For which
	\begin{equation}\label{short string approx}
	z_- \approx \frac{k^2-c_1^2}{\omega^2} ~~\text{and}~~ z_+\approx \frac{1}{\varkappa^2}~~~ \text{with~~} \omega>>k.
	\end{equation}
	Using this approximation and the periodicity condition (\ref{periodicity condition}), we get
	\begin{equation}
	\frac{k}{\omega}=\sqrt{z_- +\frac{B^2 \pi^2 (1+z_-)}{4(1+\varkappa^2-B^2\varkappa^2) \mathbf{K}^2\left(\frac{z_-(1+\varkappa^2)}{z_-+1}\right)}}.
	\end{equation}
	Substituting these above result in (\ref{spinning in ads general energy}) and (\ref{spinning in ads general spin}) and then expanding for small $z_-$, the leading order terms in energy and spin expressions for sort string  turn out to be
	\begin{eqnarray}
	%E &\approx&\frac{2\sqrt{(1+\varkappa^2)(1+z_-)}}{\alpha'\pi(1-z_-\varkappa^2)}\sqrt{z_- +\frac{B^2 \pi^2 (1+z_-)}{4(1+\varkappa^2-B^2\varkappa^2) \mathbf{K}^2\left(\frac{z_-(1+\varkappa^2)}{z_-+1}\right)}} \mathbf{E}\left(\frac{z_-(1+\varkappa^2)}{z_-+1}\right) \nonumber \\
	E&\approx&  \mathcal{A}\left(1+\frac{(1 +\varkappa^2+B^2)z_- }{2B^2}-\frac{2
		\left(1+\varkappa ^2\right)^2-12 B^2\varkappa ^2 \left(1+\varkappa ^2\right)+B^4 \left(1-6 \varkappa ^2+3 \varkappa ^4\right)}{16 B^4}z_-^2\right), \nonumber \\
	S &\approx&	 \frac{\sqrt{1+\varkappa^2}}{2 \alpha'}\left[z_- +\frac{3(-1+\varkappa^2)}{8}z_-^2\right],
	\end{eqnarray}
	with $\mathcal{A}=\frac{B\sqrt{1+\varkappa^2}}{\alpha'\sqrt{1+\varkappa^2-B^2 \varkappa^2}}$. Therefore  the expression for energy in terms of spin can be written as 
	\begin{equation}
	E^2 \approx \frac{(1+\varkappa^2)}{\alpha'^2(1+\varkappa^2-B^2 \varkappa^2)}\left(B^2+\frac{2\alpha'(1+B^2+\varkappa^2 )}{\sqrt{1+\varkappa^2}}S\right).
	\end{equation}
	For small oscillation  i.e. $B \rightarrow 0$ 
	\begin{equation}
	E^2 \approx 2 \frac{\sqrt{1+\varkappa^2}}{\alpha'}S.
	\end{equation} 
	For $\varkappa \rightarrow 0 $ the above results reduces to result obtained in undeformed case  \cite{Khan:2003sm},\cite{Frolov:2002av}.\\
	\underline {\textbf{Long string limit}} \\
	Now let us consider the long string solution, for which 
	the length of string becomes maximum i.e. $\rho_-\approx \rho^*$. Now using (\ref{roots of rho}), (\ref{accessible region of rho}) and (\ref{inequality condition }), we get
	\begin{equation}
	z_-=\sinh^2 \rho_-=\sqrt{\left(1+\frac{1}{\varkappa^2}\right)\left(1-\frac{c_1^2}{\omega^2}\right)}-1 ,
	\end{equation}
	\begin{equation}
	\frac{k}{\omega}=\sqrt{1+\varkappa^2}-\varkappa \sqrt{1-\frac{c_1^2}{\omega^2}}.
	\end{equation}
	We can also notice that for long string solution $\rho_-$ becomes very nearer to $\rho_+$. Therefore 
	taking $\frac{z_-}{z_+}=1-\varepsilon$ where $\varepsilon<<1$, the expressions for energy and spin for long string limit become
	\begin{eqnarray}
	E=\frac{2 k \sqrt{1+\varkappa^2}}{\pi \alpha'\varkappa \omega \sqrt{z_-(z_-+1)}} \sqrt{1-\varepsilon}~\mathbf{\Pi}\left(\frac{z_-(1+\varkappa^2)}{z_-+1},1-\frac{\varepsilon}{1+z_-},
	\right) \label{spinning in ads long string enery}\\
	S=  \frac{2\sqrt{ 1+\varkappa^2}}{\pi \alpha'\varkappa\sqrt{z_-(z_-+1)}}
	\sqrt{1-\varepsilon}~\left( \mathbf{\Pi}\left(\frac{z_-}{z_-+1},1-\frac{\varepsilon}{1+z_-}
	\right)- \mathbf{K
	}\left(1-\frac{\varepsilon}{1+z_-}\right).
	\right) \label{spinning inads long string spin}
	\end{eqnarray}
	Now using the expansions of elliptic integrals for $\varepsilon<<1$
	\begin{eqnarray*}
		\sqrt{1-\varepsilon}~\mathbf{\Pi}\left(m,1-\frac{\varepsilon}{1+z_-}\right)= \frac{4 \log2-\sqrt{m}\log\left[\frac{1+\sqrt{m}}{1-\sqrt{m}}\right]-\log\frac{\varepsilon}{1+z_-}}{2(1-m)} +\mathcal{O(\varepsilon)},\nonumber \\
		\sqrt{1-\varepsilon}~\left[\mathbf{\Pi}\left(n,1-\frac{\varepsilon}{1+z_-}\right) -\mathbf{K}\left(1-\frac{\varepsilon}{1+z_-}\right)\right]=\frac{n\left(4 \log2-\frac{1}{\sqrt{n}}\log\left[\frac{1+\sqrt{n}}{1-\sqrt{n}}\right]-\log\frac{\varepsilon}{1+z_-}\right)}{2(1-n)} +\mathcal{O(\varepsilon)} \nonumber\\
	\end{eqnarray*}
	and identifying $m=\frac{z_-(1+\varkappa^2)}{z_-+1}$ and $n=\frac{z_-}{z_-+1}$ , we get
	\begin{eqnarray}
	E=\frac{\sqrt{1+\varkappa^2}}{\pi \alpha'\varkappa \sqrt{m}}\left(4 \log2-\sqrt{m}\log\left[\frac{1+\sqrt{m}}{1-\sqrt{m}}\right]-\log\frac{\varepsilon}{1+z_-}\right), \\
	S=\frac{\sqrt{1+\varkappa^2}}{\pi \alpha'\varkappa }\sqrt{n}\left(4 \log2-\frac{1}{\sqrt{n}}\log\left[\frac{1+\sqrt{n}}{1-\sqrt{n}}\right]-\log\frac{\varepsilon}{1+z_-}\right),
	\end{eqnarray}
	where we have used the relations
	\begin{equation}
	\frac{k}{\omega\sqrt{z_-(z_-+1)} (1-m)}=\frac{1}{\sqrt{m}} ~~\text{and}~~  \frac{n}{\sqrt{z_-(z_-+1)} (1-n)}=\sqrt{n}.
	\end{equation}
	Now  the relation between energy and spin turns out to be
	\begin{equation}
	E-\frac{1}{\sqrt{mn}}S= \frac{\sqrt{1+\varkappa^2}}{\pi \alpha'\varkappa}\left(\log\left[\frac{1-\sqrt{m}}{1+\sqrt{m}}\right]+\frac{1}{\sqrt{mn}}\log\left[\frac{1+\sqrt{n}}{1-\sqrt{n}}\right]\right).
	\end{equation}
	finally substituting $\frac{1}{\sqrt{m}}=\omega_0$ and $\sqrt{n}=k_0$, we get
	\begin{eqnarray}
	E-\frac{\omega_0}{k_0}S=\frac{\sqrt{1+\varkappa^2}}{\pi \alpha'\varkappa}\left(\log \left[\frac{\omega_0-1}{\omega_0+1}\right]+\frac{\omega_0}{k_0}\log\left[\frac{1+k_0}{1-k_0}\right]\right).
	\end{eqnarray}
	The above relation agrees with result found in \cite{Kameyama:2014vma}.

	\section{Spinning  both in $\left(\text{ AdS}_5\right)_\varkappa$ and $\left(\text{S}^5\right)_\varkappa$}
	In this section we consider a string which spins both in $\left(\text{ AdS}_5\right)_\varkappa$ and $\left(\text{S}^5\right)_\varkappa$. The ansatz for such string can be written as
	
	\begin{equation}
	t=t(\tau), ~~~\rho= \rho(\sigma), ~~~  \psi= \frac{\pi}{2},~~~\psi_2= \omega \tau,~~~ \theta=\theta(\sigma),~~~ \varphi=\nu \tau.
	\end{equation}
	Since the embeddings of $AdS_5$ part remains unchanged from the previous section, the equation of motion of $\rho$ and so as its solution and periodicity condition are remain same. While the equation of motion $\theta$ and Virasoro constraint equation are given by 
	\begin{equation}
	2 \theta''\left(1+\varkappa^2 \cos^2 \theta \right)+\varkappa^2\theta'^2\sin 2 \theta +\nu^2\sin 2 \theta (1+\varkappa^2)=0,
	\end{equation}
	\begin{equation}
	-\frac{\cosh^2 \rho k^2 }{1-\varkappa^2 \sinh^2 \rho }+ \frac{ \rho'^2}{1-\varkappa^2 \sinh^2 \rho}+\omega^2\sinh^2\rho+ \frac{ \theta'^2}{1+\varkappa^2 \cos^2 \theta}+\frac{\nu^2\sin^2 \theta}{1+\varkappa^2 \cos^2 \theta }=0. 
	\end{equation}
	The solution to the equation motion of $\theta$ can be obtained in terms of elliptic function 
	\begin{equation}
	\sin \theta= 
	\begin{cases}
	\mathbf{sn} \left(c_1 \sqrt{1+\varkappa^2} \sigma, \frac{1}{B^2}
	\right) 		& \text{for~} B > 1 \\
	B~\mathbf{sn}\left( \sqrt{\nu^2+c_1^2 \varkappa^2} \sigma , B^2
	\right)            & \text{for~} B < 1,
	\end{cases}
	\end{equation}
	where $B=\sqrt{\frac{c_1^2(1+\varkappa^2)}{\nu^2+c_1^2 \varkappa^2}}$ and $c_1$ is an integration constant.
	There is another periodicity condition comes from the string configuration along $\theta$ angle, which gives
	%	\begin{equation}
	%	\pi \sqrt{c1^2(1+\varkappa^2)} =2 \mathbf{K}\left(\frac{1}{b}\right)
	%	\end{equation}  
	
	\begin{eqnarray}
	2\pi \sqrt{\nu^2 +\varkappa^2 c_1^2}&=&4 \mathbf{K}\left[\frac{c_1^2(1+\varkappa^2)}{\nu^2+\varkappa^2c_1^2}\right]. \nonumber 
	\end{eqnarray}
	The general expressions of  conserved charges $E$ and $S$ will be same as (\ref{spinning in ads general energy}) and (\ref{spinning in ads general spin}), while the extra angular momentum  can be evaluated as
	\begin{eqnarray}
	\mathcal{J}&=& \frac{\nu\sqrt{1+\varkappa^2}}{2 \pi \alpha'}\int_0^ {2\pi} d\sigma \frac{\sin^2 \theta}{1+\varkappa^2 \cos^2 \theta} =\frac{2\nu \sqrt{1+\varkappa^2}}{\alpha'\pi}\int_0^{\theta_{max}} \frac{d\theta}{\theta'}\frac{\sin^2 \theta}{1+\varkappa^2 \cos^2 \theta}. \nonumber \\
	\text{So~~~} \mathcal{J}&=& \frac{2 \nu \sqrt{1+\varkappa^2}}{\alpha'\pi \varkappa^2 \sqrt{\nu^2+c_1^2\varkappa^2}} \left[\mathbf{\Pi}\left(\frac{\varkappa^2 B^2}{1+\varkappa^2},B^2\right)-\mathbf{K}(B^2)\right].
	\end{eqnarray}
	Now using the periodcity condition of $\theta$ and $\rho$  we express all the parameters in terms of $B$ and $z_-$
	\begin{equation}
	c_1=\frac{2B \mathbf{K}[B^2]}{\pi \sqrt{1+\varkappa^2}},~~~~~\nu=\frac{2}{\pi}\sqrt{\frac{1+\varkappa^2(1-B^2)}{1+\varkappa^2}}\mathbf{K}[B^2],
	\end{equation}
	\begin{equation}
	\omega=\frac{2}{\pi\sqrt{1+z_-}}\mathbf{K}\left(\frac{(1+\varkappa^2)z_-}{1+z_-}\right),~~k^2=\frac{4}{\pi^2}\left[\frac{z_-}{(1+z_-)}\mathbf{K}^2\left(\frac{(1+\varkappa^2)z_-}{1+z_-}\right)+\frac{B^2}{(1+\varkappa^2)}\mathbf{K}^2(B^2)\right].
	\end{equation}
	Now substituting these relation in conserves charges we get
	\begin{eqnarray}
	E &=&\frac{2\sqrt{(1+\varkappa^2)(1+z_-)}}{\alpha'\pi(1-z_-\varkappa^2)}\frac{\sqrt{1+z_-}}{\mathbf{K}\left(\frac{z_-(1+\varkappa^2)}{z_-+1}\right)}\sqrt{\frac{B^2 \mathbf{K}^2(B^2)}{1+\varkappa^2}+\frac{z_- \mathbf{K}^2\left(\frac{z_-(1+\varkappa^2)}{z_-+1}\right)}{1+z_-}} \mathbf{E}\left(\frac{z_-(1+\varkappa^2)}{z_-+1}\right), \nonumber \\
	%&\approx& C\frac{\sqrt{1+\varkappa^2}}{\alpha'\sqrt{1+\varkappa^2-C^2 \varkappa^2}}\left(1+\frac{(1+C^2 +\varkappa^2)z_- }{2C^2}-\frac{2
	%\left(\varkappa ^2+1\right)^2-12 C^2\varkappa ^2 \left(1+\varkappa ^2\right)+C^4 \left(1-6 \varkappa ^2+3 \varkappa ^4\right)}{16 C^4}z_-^2\right) \nonumber \\
	S &=&\frac{2\sqrt{ 1+\varkappa^2}}{\pi \alpha'\sqrt{(z_-+1)}}
	\left[ \mathbf{\Pi}\left(\frac{z_-}{z_-+1},\frac{z_-(1+\varkappa^2)}{z_-+1}
	\right)- \mathbf{K
	}\left(\frac{z_-(1+\varkappa^2)}{z_-+1}\right) \right], \nonumber \\
	J&=&\frac{2\sqrt{1+\varkappa^2-\varkappa^2B^2}}{\alpha'\pi \varkappa^2}\left[\mathbf{\Pi}\left(\frac{\varkappa^2 B^2}{1+\varkappa^2},B^2\right)-\mathbf{K}(B^2)\right].
	\end{eqnarray}	
	Using the short string limit $z_-\rightarrow 0$ and $B<<1$, the leading terms in $E, S \text{~and~} \mathcal{J}$ turn out to be
	\begin{eqnarray}
	E^2 \approx \frac{1}{\alpha'^2}\left(B^2+\left(1+\varkappa^2\right)z_-\right),	\nonumber \\
	S \approx \frac{\sqrt{1+\varkappa^2}}{2\alpha'}z_-,~~~\text{and}~~~~\mathcal{J} \approx \frac{B^2}{2 \alpha' \sqrt{1+\varkappa^2}}.
	\end{eqnarray}
	Finally the relation between $E, S \text{~and~} \mathcal{J}$ becomes
	\begin{eqnarray}
	E^2\approx 2 \frac{\sqrt{1+\varkappa^2}}{\alpha'}\left(J+S\right).
	\end{eqnarray}
	For $\varkappa\rightarrow 0$, the above result reduces to result found in \cite{Russo:2002sr}. This scaling relation is of the form of usual Regge-type spectrum. In the case of $\varkappa\rightarrow 0$ the dual gauge theory operator has been identified in \cite{Russo:2002sr}. In the present case of nonzero $\varkappa$, the proper identification is still missing even for the simplest class of folded and spinning string solutions.  
	
	\section{Summary and Conclusion}
	In this paper, we have studied a generic class of semiclassical string solitons in $\varkappa$ deformed $AdS_5 \times S^5$ background. We have restricted ourselves to the bosonic part of the string action in conformal gauge. We have found the exact solutions for the string configuration which are pulsating in $(AdS_5)_{\varkappa}$ and at the same time spinning in the $S^5_{\varkappa}$ and vice versa and computed the general expressions of conserved charges in terms Jacobi elliptic functions and elliptic integrals. We have obtained the scaling relation between the energy and angular momentum for such configurations in short  and long string limits. As expected these relations reduce to their corresponding undeformed results when $\varkappa \rightarrow 0$. We have further studied an example of string solition which spins along both subspaces of $(AdS_5 \times S^5)_{\varkappa}$ and write down the scaling relation among the charges in the leading orders.\par
	There are some open questions which could be addressed later. First, it would be interesting to study the corresponding gauge theory operator, if any, dual to these string solitons. Recall 
	that the novelty of the solutions presented in \cite{Khan:2003sm} was to find out a scaling relation between various charges through an additional constant, whose gauge theory interpretation was highly nontrivial. It was argued that suitably adjusting this nontrivial constant one was able to see some direct relationship with the underlying gauge theory.
	Ours is a generalization of those results and we have been able to successfully incorporate all nontrivial factors which have been introduced by the deformation parameter $\varkappa$. These are a generic class of solutions for the string trajectories in the gravity side. We hope this will give further insight in the study of string solutions in $\varkappa$-deformed $AdS_5 \times S^5$. The recent studies suggest that the underlying gauge theory might be some nonlocal theory which at best remains obscure. Therefore, studying classical rotating and pulsating string solutions are expected to give us hints along these directions. Secondly, as the deformed geometry is question is not well understood as a generic string background it would be interesting to see if we can take clue from the deformed supergravity equations proposed in  \cite{Arutyunov:2015mqj} and look for possible ``modified'' intersecting D-branes whose near horizon geometry gives rise to these deformed geometries.  Thirdly, finding the multi spin string solution in this deformed background is worth attempting, though in this case the string rotating with equal spins along the two subspace will not be possible because of loss of symmetry due to the deformation.  Finally, one can do a one loop calculation to find out the leading order corrections to the conserved charges along the lines of \cite{Beccaria:2010zn}. We will report on some of these issues in near future.
	%\newpage
	
	%\bibliographystyle{JHEP}
	%\bibliography{ref-finalv2}

\providecommand{\href}[2]{#2}\begingroup\raggedright\endgroup

\end{document}